\documentclass[ reprint,  amsmath,amssymb,  aps, superscriptaddress]{revtex4-1}%
\usepackage{graphicx}
\usepackage{dcolumn}
\usepackage{bm}
\usepackage{amsmath}
\usepackage{amsfonts}
\usepackage{amssymb}%
\usepackage{txfonts}

\begin{document}
\title{Finite-Size Effects on K\'arm\'an Vortex in Molecular Dynamics Simulation}
\author{Yuta Asano}
\email{yuta.asano@issp.u-tokyo.ac.jp}
\affiliation{Institute for Solid State Physics, The University of Tokyo, Kashiwa, Chiba 277-8581, Japan}
\author{Hiroshi Watanabe}
\affiliation{Institute for Solid State Physics, The University of Tokyo, Kashiwa, Chiba 277-8581, Japan}
\affiliation{Department of Applied Physics and Physico-Informatics, Keio University, Yokohama, Kanagawa 223-8522, Japan}
\author{Hiroshi Noguchi}
\affiliation{Institute for Solid State Physics, The University of Tokyo, Kashiwa, Chiba 277-8581, Japan}


\begin{abstract}
The characteristics of the K\'arm\'an vortex generated by a molecular dynamics (MD) simulation exhibit strong finite-size effects, and MD can only reproduce the experimental results qualitatively. Here, we seek the simulation conditions for quantitatively reproduce the vortex shedding frequency. We found that the finite-size effects are mainly caused by the interference of vortices and the nonuniformity of the fluid viscosity coefficient.
\end{abstract}


\maketitle

It is well known that an obstacle in a flow generates two rows of staggered vortices called K\'arm\'an vortex street behind the obstacle.
Flows around an obstacle can be found in many places around us, such as airflow around a building, water flow around a screw propeller, and so forth.
Because the K\'arm\'an vortex is a source of noise and vibration, it is crucially important to understand the flow characteristics around an obstacle in the engineering field.
Especially, the flow around a circular cylinder is one of the most fundamental problems of fluid dynamics.
Therefore, experimental and numerical studies have been conducted on the flow around the circular cylinder~\cite{williamson96}.
The shedding characteristics of the vortices are phenomenologically well understood for the Newtonian fluid.
The shedding frequency is given experimentally by the following equation~\cite{roshko54}:
\begin{eqnarray}
St=0.212 - \frac{4.5}{Re}, \label{eq1}
\end{eqnarray}
where $St=fD/V$ and $Re=\rho DV/\eta$ are the Strouhal number and Reynolds number, respectively ($f$: vortex shedding frequency, $D$: cylinder diameter, $V$: inflow velocity, $\rho$: fluid density, $\eta$: fluid viscosity coefficient).

With the progress in computational power, K\'arm\'an vortex can be generated by molecular dynamics (MD) simulation~\cite{rc86, rapaport87, awn18}.
Although MD simulation is advantageous to deal with the complex flow phenomena directly, such as the Toms effect~\cite{gadd65} exhibited by polymer solutions, and cavitation in multiphase flow~\cite{gm16}, it suffers from strong finite-size effects.
To understand the finite-size effects on MD, we investigate the difference between the Strouhal number $St^{\rm (MD)}$ obtained by MD simulation and the value predicted by Eq.~(\ref{eq1}).
This relation Eq.~(\ref{eq1}) holds for not only macroscopic methods based on the Navier-Stokes equations but also mesoscale methods~\cite{hd97a, lg02}.
However, the values obtained by the MD simulations are twice as large as the expected value~\cite{rc86, rapaport87, awn18}.
Because both $St$ and $Re$ are dimensionless quantities that do not depend on a typical spatial scale such as the cylinder diameter, it is expected that the relation Eq.~(\ref{eq1}) also holds in MD simulation.
Possible causes of the deviation are the influence of compressibility~\cite{sd03, ouv06} due to high flow velocity and the interference of vortices due to periodic boundary conditions~\cite{swp99, twt14}.
The effects can be investigated by changing the density for the compressibility and the channel width for the vortex interference.
Therefore, in the present study, we investigate the condition that quantitatively reproduces Eq.~(\ref{eq1}) for the K\'arm\'an vortex behind the cylinder by the MD simulation.

We used the Weeks--Chandler--Andersen potential~\cite{wca71} $\phi=4\epsilon\left[\left(\sigma/r\right)^{12} - \left(\sigma/r\right)^{6} + 1/4 \right]\theta(2^{\frac{1}{6}}-r)$ for the interparticle interaction of the fluid,
where $\theta$ is the Heaviside function, $r$ is interparticle distance, $\epsilon$ and $\sigma$ represent the energy and the length scale, respectively.
The mass of the particle is $m$. 
Hereafter, physical quantities are expressed in units of energy $\epsilon$, length $\sigma$, and time $\tau_0=\sigma\sqrt{m/\epsilon}$.
The simulation box is almost the same as in Ref. \cite{awn18} and is a rectangle with dimension $L_x \times L_y$, where $L_x=3600$ and $500\le L_y\le5000$.
The flow direction of the fluid is the $x$-direction.
The periodic boundary condition is taken for all directions.
The circular cylinder is modeled by a set of particles whose positions are fixed on the cylinder surface.
The cylinder with diameter $D=100$ is located at $(x, y)=(1000, L_y/2)$.
To control the temperature and the inflow velocity, we used the Langevin thermostat~\cite{awn18} in the region of $3200 \le x \le 3600$.
The friction coefficient of the Langevin thermostat ($\zeta$ of Eq.~(4) in Ref.~\cite{awn18}) is $0.1$ and is linearly increased from $0.0001$ to $0.1$ at $3200 \le x \le 3400$.
The temperature is set to $T=1$.
This thermostat relaxes
the velocities of the fluid particles into the Maxwell distribution whose average velocity is given by $V$ in the $x$-direction.
The density is in the range of $0.4 \le \rho \le 0.83$.
In the three-dimensional (3D) simulations,
the thickness of the simulation box in the $z$-direction $L_z=10$ is employed.
When K\'arm\'an vortices appear, a periodic lifting force $F_{\rm L}$ whose period is equal that of the vortex shedding acts on the cylinder.
Here, we adopted the frequency of lift coefficient $C_{\rm L}=2F_{\rm L}/(\rho D V^2)$ as the shedding frequency of the K\'arm\'an vortex.
MD simulations were performed using the velocity-Verlet algorithm with a time step of $0.004$, for up to a maximum of $10~000~000$ time steps.
The time integration is performed by LAMMPS~\cite{plimpton95}.

\begin{figure}
\includegraphics{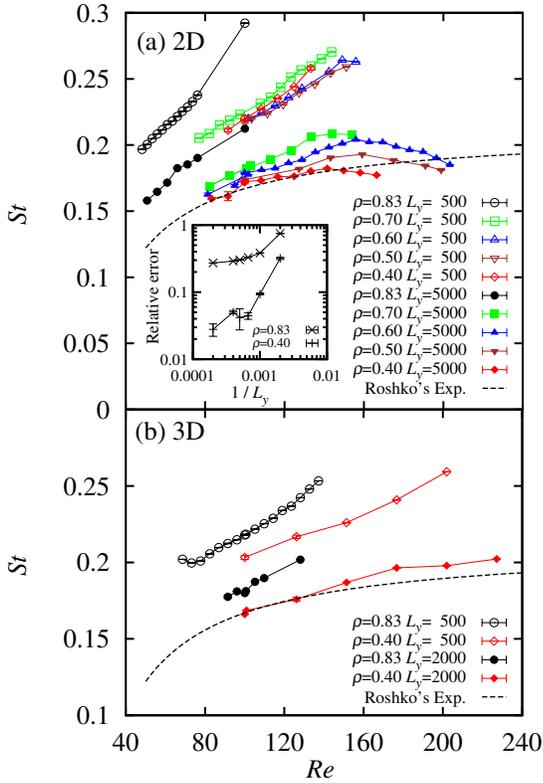}
\caption{(Color online) Strouhal number $St$ as a function of the Reynolds number $Re$ in (a) 2D and (b) 3D.
The open and filled symbols show the results of the channel width $L_y$=500 and $5000$, respectively. The inset in panel (a) shows the relative error at $Re=100$ as a function of $L_y$.}
\label{f1}
\end{figure}

Figure~\ref{f1}(a) shows the $Re$ dependence of $St^{\rm(MD)}$ in the two-dimensional (2D) simulations.
The broken line in the figure shows Eq.~(\ref{eq1}).
While all results deviate from Eq.~(\ref{eq1}) in the case of $L_y=500$,
the results of $L_y=5000$ tend to approach Eq.~(\ref{eq1}).
Because we controlled $Re$ by changing $V$, the Mach number $Ma$ increases with increasing $Re$.
The decreasing in $St$ of $L_y=5000$ at high Reynolds number ($Re \gtrsim 160$) is due to the high Mach number $Ma$. Because $Ma$ is 0.5 or more in the range of $Re \gtrsim 160$, the compressibility affects the vortex shedding frequency. The compressibility effects on the frequency also appear in the simulation based on the Navier-Stokes equations~\cite{sd03, ouv06}. In addition, because the flow velocity is too fast in the MD simulation, the formation of vortices is inhibited due to the detachment of the fluid behind the cylinder.
The inset of Fig.~\ref{f1} shows the $L_y$ dependence of the relative error of $\rho=0.40, 0.83$ at $Re=100$.
The relative error is defined as $\left|St^{\rm (MD)}-St\right|/St$.
In the case of $\rho=0.40$, the relative error decreases monotonically as $L_y$ increases.
Therefore, the interference of the vortex with the image cell is one of the main causes of the deviation from Eq.~(\ref{eq1}).
As for $\rho=0.83$, the relative error only reaches up to $\sim 30\%$ and the relative errors are almost unchanged between $L_y=2500$ and $L_y=5000$.
This observation suggests that another cause also exists for this phenomenon at the high density.
Figure~\ref{f1}(b) shows the results of the 3D simulations.
Like the 2D simulations, $St^{\rm(MD)}$ tends to approach Eq.~(\ref{eq1}) as $L_y$ increases.

\begin{figure}
\includegraphics{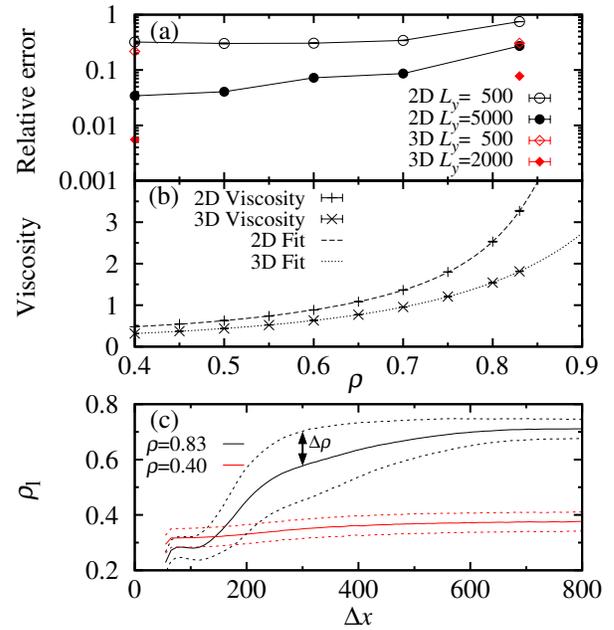}
\caption{(Color online) (a) Relative error at $Re=100$ and (b) the viscosity as a function of the density. The open and filled symbols show the results of the channel width $L_y$=500 and $L_y=5000$, respectively. The dashed and dotted lines in (b) are obtained by fits using Eq.~(19) in Ref.~(\cite{hdt13}) with $d=4$. (c) Time-averaged local density $\rho_{\rm l}$ as a function of the downstream distance $\Delta x$ at $Re=100$ with $L_y=5000$ in 2D case. Solid lines show the average, and dotted lines show upper and lower limits of fluctuations.}
\label{f2}
\end{figure}

The density dependence of the relative error at $Re=100$ is shown in Fig.~\ref{f2}(a).
As the density decreases, the error at $L_y=5000$ decreases monotonically
while it is almost independent at $L_y=500$ for the low density ($\rho \le 0.6$).
Therefore, the deviation from Eq.~(\ref{eq1}) decreases as the density decreases when the system is almost free from the interference of the vortices.
As shown in Fig.~\ref{f2}(b), the viscosity increases rapidly with the density.
Therefore, the variation in the viscosity due to the density change also increases with increasing the density.
In flows, the spatial and temporal density variation is caused by the pressure change.
Figure~\ref{f2}(c) shows the time-averaged local density $\rho_{\rm l}$ and the temporal density fluctuation $\Delta \rho$ downstream of the cylinder at Re = 100 with $L_y=5000$ in the 2D case. A characteristic change in $\Delta\rho$ exists at $200<\Delta x <300$ in the case of $\rho = 0.83$. The maximum density fluctuation of $\rho = 0.40$ and $\rho = 0.83$ are $\Delta\rho = 0.035$ and $\Delta\rho = 0.12$, respectively, where $\Delta x$ is the downstream distance from the center of the cylinder.
Since $Ma$ is not so different ($Ma\simeq0.4$ and $0.5$ at $\rho=0.40$ and $0.83$, respectively), it is not caused by $Ma$ change.
The average density behind the cylinder is spatially lower than the other regions at higher density. Since these density variations induce nonuniform viscosity, the effective Reynolds number which determines the vortex shedding frequency is altered.
Hence, we conclude that the nonuniformity of viscosity near the cylinder is the cause of $St$ deviation at high density.
The relative errors in 3D are smaller than those in 2D because the viscosity variation is smaller as shown in Fig.~\ref{f2}.

In summary, we evaluate the Strouhal number of a K\'arm\'an vortex behind a circular cylinder by MD simulations.
The main causes of the deviation from the experimental results are the interference of vortices and the nonuniformity of the viscosity.
The former can be suppressed by widening the channel width, and the latter can be reduced by decreasing density.
Therefore, we conclude that a nanoscale K\'arm\'an vortex can be analyzed quantitatively by the MD simulation.
\begin{acknowledgments}
This research was supported by MEXT as ``Exploratory Challenge on Post-K computer'' (Challenge of Basic Science—Exploring Extremes through Multi-Physics and Multi-Scale Simulations) and JSPS KAKENHI Grant No.~JP15K05201.
Computation was partially carried out by using the facilities of the Supercomputer Center, Institute for Solid State Physics, University of Tokyo.
\end{acknowledgments}


\begin{thebibliography}{9}
\bibitem{williamson96}C. H. K. Williamson, Annu. Rev. Fluid Mech. {\bf 28}, 477 (1996). 
\bibitem{roshko54}A. Roshko, NACA Report 1191 (1954), Printed in USA.
\bibitem{rc86}R. C. Rapaport and E. Clementi, Phys. Rev. Lett. {\bf 57}, 695 (1986).
\bibitem{rapaport87}D. C. Rapaport, Phys. Rev. A {\bf 36}, 3288 (1987).
\bibitem{awn18}Y. Asano, H. Watanabe, and H. Noguchi, J. Chem. Phys. {\bf 148}, 144901 (2018).
\bibitem{gadd65}G. E. Gadd, Nature {\bf 206}, 463 (1965). 
\bibitem{gm16}A. Gnanaskandan, and K. Mahesh, J. Fluid. Mech. {\bf 790}, 453 (2016).
\bibitem{hd97a}X. He and G. Doolen, J. Comput. Phys. {\bf 134}, 306 (1997).
\bibitem{lg02}A. Lamura and G. Gompper, Eur. Phys. J. E {\bf 9}, 477 (2002)
\bibitem{sd03} M. Sabanca and F. Durst, Phys. Fluids {\bf 15}, 1821 (2003).
\bibitem{ouv06} A. I. Osipov, A. V. Uvarov, and N. A. Vinnichenko, Phys. Fluids {\bf 18}, 105106 (2006).
\bibitem{swp99}D. Sumner, S. S. T. Wong, S. J. Price, and M. P. Paidoussis, J. Fluid. Struct. {\bf 13}, 309 (1999).
\bibitem{twt14} Z. Tr\'{a}vn\'{i}\v{c}ek, A.-B. Wang, and W.-Y. Tu, Exp. Fluids {\bf 55}, 1679 (2014).
\bibitem{wca71} J. D. Weeks, D. Chandler, and H. C. Andersen, J. Chem. Phys. {\bf 54}, 5237 (1971). 
\bibitem{plimpton95}S. J. Plimpton, J. Comput. Phys. 117, 1 (1995).
\bibitem{hdt13}R. Hartkamp, P. J. Daivis, and B. D. Todd, Phys. Rev. E {\bf 87}, 032155 (2013).
\end{thebibliography}
\end{document}